\documentclass{article}

\usepackage{arxiv}
\usepackage{graphicx}
\usepackage{dirtytalk}
\pdfoutput=1
\graphicspath{ {./img/} }
\usepackage[utf8]{inputenc} 
\usepackage[T1]{fontenc}    
\usepackage{hyperref}       
\usepackage{url}            
\usepackage{booktabs}       
\usepackage{amsfonts}       
\usepackage{nicefrac}       
\usepackage{microtype}      
\usepackage{lipsum}

\title{Ants-Review: a Protocol for Incentivized \\ Open Peer-Reviews on Ethereum}

\author{
  Bianca Trovò\thanks{Atomic Energy and Alternative Energies Commission (CEA), Frédéric Joliot Institute for Life Sciences, NeuroSpin, Cognitive Neuroimaging Unit, Saclay, France.} \\
  Sorbonne University\\
  Faculty of Science and Engineering\\
  Paris, France\\
  \texttt{bianca.trovo@protonmail.com} \\
   \And
  Nazzareno Massari\\
  MakerDAO\\
  Community Development\\
  London, United Kingdom\\
  \texttt{nazzareno@nazzarenomassari.com} \\
}

\begin{document}
\maketitle

\begin{abstract}
Peer-review is a necessary and essential quality control step for scientific publications but lacks proper incentives. Indeed, the process, which is very costly in terms of time and intellectual investment, not only is not remunerated by the journals but is also not openly recognized by the academic community as a relevant scientific output for a researcher. Therefore, scientific dissemination is affected in timeliness, quality and fairness. Here, to solve this issue, we propose a blockchain-based incentive system that rewards scientists for peer-reviewing other scientists’ work and that builds up trust and reputation. We designed a privacy-oriented protocol of smart contracts called Ants-Review that allows authors to issue a bounty for open anonymous peer-reviews on Ethereum. If requirements are met, peer reviews will be accepted and paid by the approver proportionally to their assessed quality. To promote ethical behaviour and inclusiveness the system implements a gamified mechanism that allows the whole community to evaluate the peer-reviews and vote for the best ones.
\end{abstract}


\section{Introduction}
Since the birth of Bitcoin \cite{BitcoinSatoshi} in 2008 as a peer-to-peer electronic cash system, blockchain technologies have spread far beyond the sole cryptocurrency domain, in particular after the implementation of general purpose smart contracts introduced by Ethereum \cite{Ethereum-Wood}.
Besides a growing number of applications ranging from De-Fi, healthcare, music industry, government, identity, to cite but a few, blockchain technology has recently started to catalyse the attention of the scientific community as well \cite{Bitcoin-Nature-focus,vanRossum2017-DigSci} with the promising potential of being a `game changer' in outdated and broken scientific practices and leading towards open science \cite{AES}. Indeed, scholars have pointed out how the intrinsic characteristics of blockchain technology set the basis for a open science infrastructure \cite{ReviewBlockchain2019} in which decisional processes are transparent and therefore more democratically accessible to all the stakeholders (researchers, reviewers, funders, taxpayers). Those are: the consensus algorithm \cite{ConsAlg}, a deterministic computational trust that allows for \emph{decentralization}, for which there are no trusted third parties; the proof of existence (PoE) that via \emph{cryptographic hashing} and \emph{timestamping} creates a digital footprint able to keep a traceable chronological record of research objects that cannot be altered or retrieved (due to its property of \emph{immutability} or \emph{append-only}) \cite{ReviewBlockchain2019}. In particular, a `blockchainified science' \cite{BlockchainforScience} could `reduce waste' \cite{ReducingWaste-Lancet}, by disclosing each step in the research cycle to `scientific self-correction' way before the final publication step, and therefore help fixing the current reproducibility crisis in science.
\newline A thorny issue in the academic system that can - and we think it should - be tackled by blockchain concerns the status and accreditation of peer review, the core process of scientific validation currently facing a crisis \cite{Gropp-PeerRevStress}.
In this paper we propose a solution to the problem of reviewers recognition based on the principles of tokenomics \cite{TokenEcon} and in line with the values of open science.

\section{Background}
\subsection{Peer review: present problems and mild solutions}
Peer review is still the only quality control mechanism devoted to evaluating scientific outcomes. The purpose of peer review is, to cite \cite{Gropp-PeerRevStress}: \say{improving the quality of the published paper, determining the originality of the manuscript, determining the importance of the findings, detecting fraud, and detecting plagiarism.}. However, the system is `ﬂawed' and outdated \cite{Smith2006} and presents multifaceted issues \cite{Walker2015-trends}, here reviewed.
\subsubsection{A slow multi-stage process.} The main issues affecting the effectiveness of peer review is the delay between paper submission and journal acceptance for publication. The traditional peer review process is centralized around the journals' editor(s). The author(s) submit the manuscript to the journal where an editorial team assesses if the paper meets the scopes of the journal and novelty criteria. If the editorial decision is to send the manuscript for review, the handling editor personally selects potential reviewers. The authors' identity is usually known to the reviewer but the reviewers' identity is hidden to the authors or among the other reviewers themselves (\emph{single-blind review}). Reviewers independently conduct their reviews by exposing in their reports strengths and weaknesses of the manuscript and sometimes substantially improving the draft. Depending on if the decision is a major or minor revision, authors are invited to re-submit a corrected or improved version of the manuscript. The same reviewers might be contacted again to continue the same peer review process. This process can take multiple rounds and is a huge time investment both for authors and reviewers. An analysis of all papers published in \emph{PubMed} for a time period of 30 years claims that the median review time is around 100 days \cite{Kendall-peerrev}.
\subsubsection{Lack of recognition and incentives.} Peer reviewing is an invisible activity purely conducted on voluntary basis, neither paid by the journals or officially credited via standard scientific metrics (such as the ones that establish the Impact Factor of an author). Thus, it does not lead to advancements in career or help securing grants. Researchers are motivated to do peer-review by a sense of belonging and a desire to `give back' to the community \cite{Warne-RewRev}. A major consequence of not promoting incentives for the quality (and quantity) of peer reviews is to either slow down publication of potentially good research which awaits for validation \cite{HauserIncent} or let bad science be published through sloppy and uncritical reviews.
\subsubsection{Fraud and misconduct.} Due to the `publish or perish culture' pressures, unethical behaviour from reviewers has been occasionally reported, from abusive behaviour towards authors \cite{Smith2006,tragedy-reviewers} to identity fraud. Some studies have reported an improvement in the transparency and civility of the review process when open reports are released according to the standards of \emph{open peer review} \cite{PeerRev-NatComm}.
\subsubsection{Social and cognitive biases.} Given the fact that anonymity is usually asymmetrically applied only for reviewers, many power related dynamics can influence the reviewers decision \cite{Tennant2017-F1000R}, such as gender or cultural discrimination and social prestige of the institution. To solve this problem some journals have implemented \emph{double-blind review} process (the identity of both authors and reviewers are masked) which seems to reduce the bias towards minorities.
\subsubsection{Peer reviews need to be... reviewed.} There is high variability in the reliability and depth of reviews and a recurrent question is: \say{Who watches the watchers?} \cite{Tennant2017-F1000R}.
\subsubsection{Need for more reviewers.} There is a disproportion between the progressive increase in journal publications and the number of experienced reviewers selected for the task which demands an expansion of the reviewer's pool including early career scientists \cite{tragedy-reviewers,Warne-RewRev}.
\\
Some mild attempts to credit peer review have been handled without much success by journals via attribution of virtual `badges', certificates of performance, citation in annual editorials \cite{Tennant2017-F1000R} where performance, though, is assessed only in terms of quantity of reviews but not quality \cite{Warne-RewRev}.
\newline Partnering with publishers, the startup \emph{Publons}\footnote[1]{\emph{Publons}. \url{https://publons.com}} provides a free metric service for tracking, verification and recognition of publications, peer reviews and journal editing in a single researcher identifier that showcases a record of scientific activity and impact based on authors' productivity.

\section{System concept}
In this paper we propose Ants-Review, a new incentivisation mechanism built on Ethereum that issues open peer reviews to validate scientific papers while preserving the anonymity of its contributors. We imagine a final paper originating from the peer review process as a complex system that emerges from the interactions between the authors and the reviewers, a whole that is more than the sum of its parts. Therefore, the name evokes an ant colony as a self-organising organism in which all micro-contributions of the individuals emerge into complex behaviour. The original proposal behind this paper can be found here \cite{AntsReview}. Its design and implementation are exposed in the following section.

\subsection{Design}

\subsubsection{Incentivization and recognition.} A popular incentive model for open source software (OSS) is represented by bounties. Bounties are prizes or monetary rewards given for completing a task before a deadline \cite{BountyGit}. Examples of such platforms that allow funders (bounty backers) to pay developers (bounty hunters) for open source contributions are \emph{Gitcoin}\footnote[2]{\emph{Gitcoin}. \url{https://gitcoin.co}} and \emph{The Bounties Network}\footnote[3]{\emph{The Bounties Network}. \url{https://bounties.network}}. Incentives can be represented by tokens, units of values registered on the blockchain. In the network of the scientific community reviewers provide a service and those who consume it (authors, journals) should be able to contribute with tokens. The amount of tokens reflects material and symbolic recognition of the performed work that can be statistically quantified for author-level metrics measuring the productivity and impact of a researcher. Thus, the system acts also as a reputation builder.

\subsubsection{Transparency and re-usability of the records.} The peer review history, including reviewers' recommendations and authors' replies, should be openly and permanently accessible to the community (in the form of `open reports' of open peer reviews) even before articles' publication in order to  make editorial decisions more democratic and prevent waste of knowledge. Following the example of models offered by journals peer review consortia, such as the \emph{Neuroscience Peer Review Consortium}\footnote[4]{\emph{Neuroscience Peer Review Consortium}. \url{http://nprc.incf.org}} and independent companies like \emph{ResearchSquare}\footnote[5]{\emph{ResearchSquare}. \url{https://www.researchsquare.com}} and \emph{Peerage of Science}\footnote[6]{\emph{Peerage of Science}.  \url{https://www.peerageofscience.org}}, that provide a scientific peer review service, peer reviews in Ants-Review will be transferable across journals (like in `cascading' or `portable peer reviews').

\subsubsection{Accountability via pseudo-anonymity.} In order to counteract malicious behaviour (see 2.1) affecting the integrity of the reviews but also to correctly attribute the intellectual contributions making sure there are no conflicts of interest, it is important to be able to track back the identities of the contributors to a peer-review report. This is possible if the platform acts like a version control system where commits are permanent and their hashes timestamped. InterPlanetary File System (IPFS) \cite{IPFS} is a peer-to-peer hypermedia protocol for storing data in a distributed file system over the internet which guarantees data immutability and unique file identification via cryptographic hashes. IPFS' hashes are then stored inside the smart contracts' state that is timestamped into the Ethereum blocks where  the transactions take place; there, data remains unaltered and indelible. This notarization process is called proof of existence (PoE) and allows manual verification of the existence of the document.
\newline To prevent retaliation for negative peer-reviews and to promote the participation of early researchers who might feel intimidated to judge the scientific work of senior authors, the Ants-Review system maintains the privacy of both authors and reviewers in a double-blind approach via Ethereum’s externally owned accounts (EOA) addresses and zero-knowledge proof (ZKP), a cryptographic method where a party can prove to another party the possession of certain information, like a secret key, without revealing that information (see 3.2).

\subsubsection{Inclusiveness via gamification.} As a final step we propose that all the community is involved in the process of peer reviewing by abolishing the editorial selection process through `open participation' (or `open interaction', `open platform' \cite{OPR-Ross-Hellauer,RossHellauer-OPR}). In this way, the pool of reviewers is enriched and allows younger researchers to get the appropriate training through interactive feedback. Moreover, peer reviews could be evaluated, commented, criticised by the other members of the scientific community, enabling a virtuous loop of verification. An interesting addition would be to introduce a rank of peer reviews resulting from the community feedback via a voting process (see 4.1). It is conceivable that the community members that engage in assessing the quality of peer reviews could be incentivized as well. This solution would create a self-reinforcing ethical behaviour where the fair evaluation of peer reviews would be also in the interest of the agents at play.

\subsection{Implementation}
\begin{figure}[htp]
\centering
\includegraphics[scale=0.28]{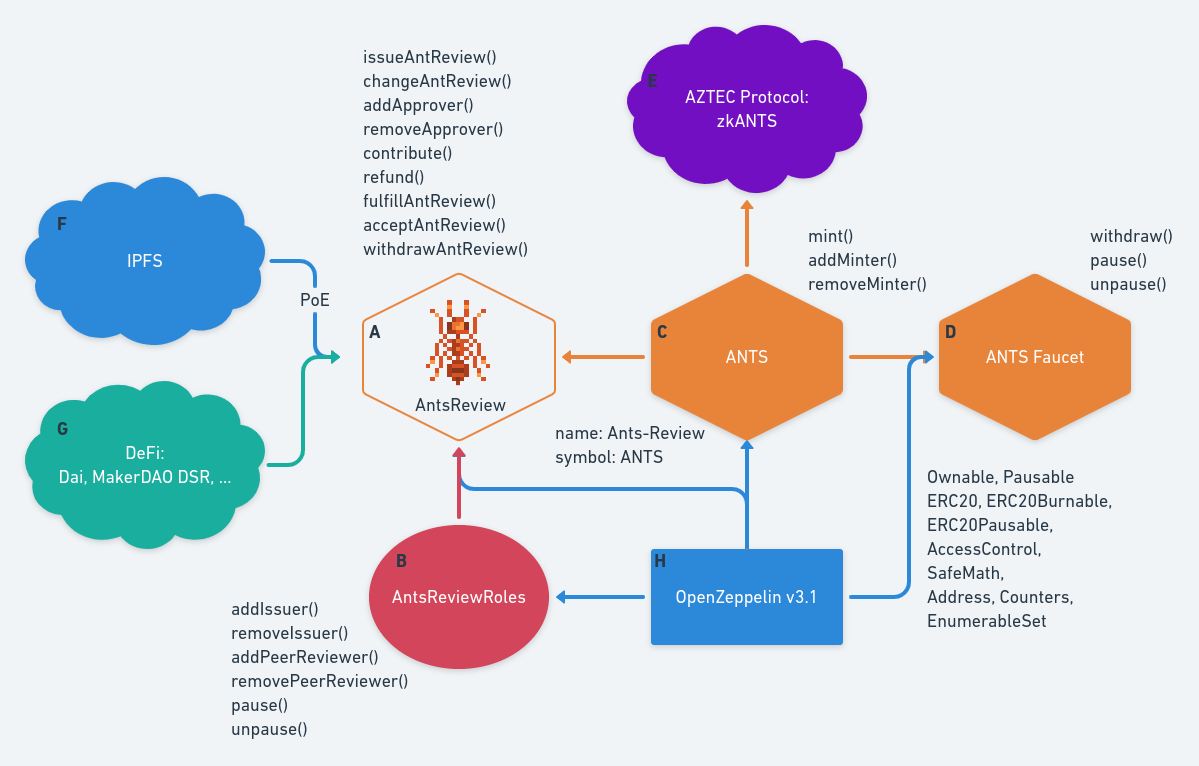}
\caption{Exagones represent the protocol's smart contracts. Ellipse represents the smart contract inherited by AntsReview. Clouds represent integrations into the protocol. Rectangles represent the smart contracts' libraries.
\textbf{a} Core contract of the protocol implementing a bounty system with the functions listed.
\textbf{b} Module inherited by AntsReview: it manages the access control of the protocol by adding and removing issuers and peer-reviewers.
\textbf{c} Native token used in the protocol. It is linked to (\textbf{a}), (\textbf{e}) and to (\textbf{d}).
\textbf{d} Faucet to distribute \emph{ANTS} on Kovan Testnet for testing purposes. 
\textbf{e} Integration with Aztec Protocol to wrap (\textbf{c}) into \emph{zkANTS} to implement private \emph{ANTS} transactions on Ethereum.
\textbf{f} Integration with IPFS to upload papers, requirements and peer reviews and store the hash as \emph{PoE} into AntsReview (\textbf{a}).
\textbf{g} Integration with ERC20 tokens like \emph{Dai}, and De-Fi services like MakerDAO DSR to be used in the protocol.
\textbf{h} Library used by the protocol for secure contract development with the modules listed.}
\label{fig:contracts}
\end{figure}

The Ants-Review Protocol is divided into different modules responsible for the following functionalities, as shown in the flow-chart (see Figure \ref{fig:contracts}):
\emph{AntsReview}, which manages access management and the core system (see Figure \ref{fig:contracts}, (\textbf{a, b, f, h}));
\emph{Privacy}, which maintains the anonymity of the system via AZTEC Protocol (see Figure  \ref{fig:contracts}, (\textbf{e}));
\emph{Tokenomics}, which manages the incentive mechanism of the system (see Figure \ref{fig:contracts}, (\textbf{c, d, e, h})). \newline Agents in the platform are: issuers, peer reviewers, approvers and contributors (or Anters, members of the Ants-Review Community).

\subsubsection{AntsReview.}
AntsReview (see Figure \ref{fig:contracts} (\textbf{a})), the core of the smart contracts written in \emph{Solidity}\footnote[7]{\emph{Solidity}. \url{https://solidity.readthedocs.io}}, a contract-oriented programming language for writing smart contracts that run on the Ethereum Virtual Machine (EVM), is deployed on Ethereum Kovan Testnet\footnote[8]{\emph{AntsReview}. \url{https://kovan.etherscan.io/address/0x85be8F04482cBB920550d5469E4dEdD6e1788121}}. AntsReview implements a bounty-like system (based on the \emph{StandardBounties} contract\footnote[9]{\emph{StandardBounties.sol}. \url{https://github.com/Bounties-Network/StandardBounties}}) where Alice (issuer) can issue an AntReview with the function \emph{issueAntReview()}.
\newline In order to create a transparent and openly accessible AntReview, Alice has to complete a series of required tasks:
 \begin{itemize}
     \item \emph{upload of the files} containing the requirements of the peer review and the paper to be reviewed into IPFS, whose hash is stored into the Ethereum blockchain as PoE;
     \item \emph{specification of a deadline} in the form of a UNIX timestamp after which the fulfillment will no longer be accepted;
     \item \emph{specification of the Issuers and an Approver} to respectively modify the AntReview and approve the peer reviews sent by the peer-reviewers.
\end{itemize}
Alice or the issuers can at any time update the AntReview details (issuers, paper, requirements, deadline) with the function \emph{changeAntReview()} and add/remove the approver with the functions \emph{addApprover()} and \emph{removeApprover()}.
 Anters (contributors) can contribute to Alice's AntReview with the function \emph{contribute()}, by specifying the amount of ANTS they are willing to send. 
Bob (peer-reviewer) can download the files relative to Alice's paper and the requirements of the peer review by leveraging on the content-addressing feature of IPFS that allows anyone to find the document using an IPFS explorer; subsequently, Bob can submit a peer review before the deadline by fulfilling the AntReview created by Alice, with the function \emph{fulfillAntReview()}, by uploading the peer review on IPFS, whose hash is stored into Ethereum blockchain as PoE. He can update the peer review with the function \emph{updateReview()} by uploading the new version on IPFS. 
Ted (approver) can accept the peer review submitted by Bob with the function \emph{acceptAntReview()}, by specifying the amount of ANTS that will be transferred as reward from the contract to Bob.
If Alice's AntReview does not receive any peer review and the deadline expires, Anters can get a refund with the function \emph{refund()} for their contributions. In order to avoid residual balance, Alice can withdraw ANTS from the AntReview's balance, if the deadline expires, with the function \emph{withdrawAntReview()}, and the contract will transfer the amount specified to Alice and update the balance.
\\
Access management of the Ants-Review protocol is defined and controlled by \emph{AntsReviewRoles} (see Figure \ref{fig:contracts} (\textbf{b})), implemented by leveraging on \emph{AccessControl.sol} by OpenZeppelin Library\footnote[10]{\emph{OpenZeppelin Library}.\textsc{URL:} \url{https://openzeppelin.com}} that is used to define the Issuer and Peer-Reviewer Roles. AntsReviewRoles also integrates a circuit breaker design pattern via \emph{Pausable.sol} by OpenZeppelin to allow the Pauser Role, granted by default to the owner of the smart contracts, to pause (or unpause) all the functions in case of a security emergency, such as an attack to the smart contracts.

\subsubsection{Privacy.}
The anonymity of an agent in the system is achieved in two ways: via pseudo-anonymity, granted through Ethereum's EOAs that can pseudo-obscure the identity of the agent, and via private transactions allowed by AZTEC \cite{AZTEC} Protocol's security layer.

\paragraph{\textbf{Pseudo-anonymity.}}EOAs in Ethereum are controlled via private keys. However, the privacy is limited by the fact that both the blockchain and its transactions are public.  Therefore, the details of the transactions are visible to anyone by browsing a block explorer (such as Etherscan\footnote[11]{\emph{Etherscan}. \url{https://etherscan.io/}}) and are subject to data mining that could extract value and identify users in the blockchain.
\paragraph{\textbf{Private transactions.}}AZTEC Protocol was conceived to enable privacy on public blockchains. It uses zero-knowledge succinct non-interactive argument of knowledge (zk-SNARKs) \cite{zSNARK} and homomorphic encryption \cite{HomEncr} to validate encrypted transactions. zk-SNARKs are ZKP that require no interaction between prover and verifier; they are used inside the Ants-Review protocol via the zkANTS token to allow private transactions between the agents. Future developments will allow to leverage on permutations over Lagrange-bases for oecumenical  noninteractive arguments of knowledge (PLONK) \cite{PLONK}, 
a universal zk-SNARK construction that reduces gas costs and improves scalability.

\subsubsection{Tokenomics.}
Ants-Review integrates a few \emph{ERC20}\footnote[12]{\emph{EIP 20}. \url{https://eips.ethereum.org/EIPS/eip-20}} tokens, each of whom plays an integral role in the functioning and anonymity of the decentralized protocol. ANTS (see Figure \ref{fig:contracts}, (\textbf{c})) is the primary protocol token and can be staked into an AntReview. It is implemented by inheriting \emph{ERC20.sol} from OpenZeppelin Library with name \emph{Ants-Review} and symbol \emph{ANTS}.
A Faucet (see Figure \ref{fig:contracts}, (\textbf{d})) is implemented to distribute ANTS token on Kovan Testnet for testing purposes. zkANTS (see Figure \ref{fig:contracts}, (\textbf{e})) is a wrapper of ANTS that will be used inside the protocol to allow for private transactions among the agents of the protocol, preserving their anonymity as well as the amount of the AntReview reward and the contributions by the Anters. It will be implemented via AZTEC Protocol \cite{AZTEC}, that uses a cryptographic engine, \emph{ACE.sol}, a contract responsible for validating the set of AZTEC ZKPs and performing any transfer instructions involving AZTEC notes, minted into a zkAsset, that can be converted into ERC20 tokens. In order to implement a zkAsset called zkANTS, \emph{zkAsset.sol}, a contract implementation of a confidential token that follows the EIP-1724 standard \footnote[13]{\emph{EIP  1724}. \url{https://github.com/ethereum/EIPs/issues/1724}} will be used as a template to build an AZTEC-compatible asset.\\
The current state of the art of Ants-Review is represented by version 0.2.0 (MVP) live on Ethereum Kovan Testnet (see Supplementary Material).

\section{Discussions}
We have described how Ants-Review protocol can solve the limitations of the current peer review system (see 2.1). In particular, the lack of recognition, the lack of transparency, fraud and misconduct can be solved via the Ants-Review Protocol (see §AntsReview in 3.2); the social and cognitive biases can be counteracted via anonymity granted by AZTEC Protocol (see §Privacy in 3.2); the slowness of the process, the need for evaluation of the peer reviews themselves and the need for increasing the number of reviewers can be worked out through the creation of the community of Anters. 
\subsection{Future developments}
An interesting aspect of the protocol is the double function of an AntReview respectively as a bounty and as a pool to stake ERC20 tokens like Dai\footnote[14]{\emph{Dai}. \url{https://docs.makerdao.com/smart-contract-modules/dai-module/dai-detailed-documentation}}. Moreover, the duration of peer-reviews consents to connect the protocol to De-Fi services with the possibility for the community to accrue interest over time via MakerDAO Dai Saving Rate (Pot\footnote[15]{\emph{Pot}. \url{https://docs.makerdao.com/smart-contract-modules/rates-module/pot-detailed-documentation}.}) or Compound\footnote[16]{\emph{Compound}. \url{https://compound.finance}}, to cite a few.
Therefore, an ERC20 \emph{pool token} would be automatically released by the protocol, representing the accrued interest on the Anters' stake over time that can be traded, sold, or held as the Anter desires.
A Decentralized Autonomous Organization (DAO) \cite{Wang-DAO,DAOContr} could be formed in the future to allow ANTS stakers to participate in the governance of important aspects of the protocol, from smart contracts upgrades to minor changes in settings across the protocol.
\newline Finally, a protocol upgrade inspired by \emph{Discover}\footnote[17]{\emph{Discover.sol}. \url{https://github.com/dap-ps/discover/blob/master/contracts/Discover.sol}}, a Web3 browser by \emph{Status}\footnote[18]{\emph{Status}. \url{https://status.im/}} and still under investigation would allow Anters to validate peer reviews via an upvote/downvote system that will consent the protocol to automatically pay out the reward to the reviewers based on the votes associated with their peer reviews.

\subsection{A new community-driven standard?}
As Tennant points out \cite{Tennant2017-F1000R}, a change is already happening in the publishing industry, especially with new born publishers opening up the review process (\emph{BioMed Central}, \emph{ELife}, \emph{Frontiers}, \emph{PeerJ}, \emph{F1000 Research}). Recently, pre-print servers, such as \emph{arXiv} and \emph{biorXiv}, started integrating peer-review services into their platforms: \emph{PREreview}\footnote[19]{\emph{PREreview.} \url{https://www.prereview.org}}, \emph{PeerCommunityIn}\footnote[20]{\emph{PeerCommunityIn} \url{https://peercommunityin.org}}, \emph{Review Commons}\footnote[21]{\emph{Review Commons.} \url{https://www.reviewcommons.org}}, \emph{PrePrint Review}\footnote[22]{\emph{PrePrint Review.} \url{https://elifesci.org/preprint-review}} and the previously mentioned \emph{Peerage of Science}. This dissociation of initial scientific dissemination and scientific validation will force the publishing industry to adapt in order to keep up with the higher quality scientific process offered by those alternative peer review platforms and justify their added value \cite{Tennant2017-F1000R}. In our proposal we also decoupled the peer review process from the publishers giving it back to the scientific community and applying incentives from tokenomics. We foresee that the future will evolve towards community-driven peer reviews: peer reviews will be more and more independent from publishers \cite{DecoupJ}, and researchers will be the ones seeking the papers to review to build reputation within the community and not journals.
\newline Enlarging the pool of reviewers to potentially an entire scientific community and accelerating the whole process requires a standard for peer reviews \cite{UPR}: for example some aspects might be taken over by AI assistants (such as the \emph{Artificial Intelligence Review Assistant (AIRA)} \cite{AIRA} leaving to the reviewers the sole task of evaluating the content of a paper. Building smart contracts for peer reviews might accelerate this novel process of standardization. We hope that soon the value of peer review as a public good will be recognized by research funders and hiring committees.
\small{
\section{Conclusion}
In this paper we addressed a crucial problem within scholarly academic communication: the peer review process. We have shown how blockchain technology could provide an efficient and viable solution to open up possible directions for a paradigm shift in scientific communication. We proposed an incentive mechanism that could solve the problems of lack of acknowledgment and trust during peer reviews. We exposed the architecture of our project for which we adopted cutting-edge tools from the open source blockchain ecosystem.
\subsection*{Supplementary Material}
Source code: \url{10.5281/zenodo.3971044}; DApp: \url{https://ants-review.on.fleek.co}.
\subsection*{Declaration of Competing Interest}
This work was mainly developed during the ETHTurin Hackathon 2020 and the Gitcoin Kernel Fellowship program 2020. These organizations, as well as the authors' affiliations, had no financial role in the design and implementation of the protocol.
\subsection*{Authors Contribution Statement}
\underline{Bianca Trovò}: Conceptualization (lead); Funding Acquisition (lead); Investigation (lead); Methodology (equal); Project Administration (equal); Supervision (equal); Validation (supporting); Visualization (supporting); Writing – original draft (lead); Writing – review and editing (equal).
\underline{Nazzareno Massari}: Investigation (supporting); Methodology (equal); Project Administration (equal); Resources (lead); Software (lead); Validation (lead); Supervision (equal); Visualization (lead); Writing – original draft (supporting); Writing – review and editing (equal).

\subsection*{Acknowledgements} We would like to thank Andy Tudhope, Mark Beylin, Matteo A. Tambussi, Evan C. Harris and the four external anonymous peer reviewers for useful comments and revisions; the FDAPP 2020 workshop chairs and speakers for questions and feedback; Mitrasish Mukherjee for contributions on the interface.}

\bibliographystyle{unsrt}  
\bibliography{antsreview}  

\end{document}